\pdfoutput=1
\documentclass[10pt]{iopart}
\usepackage{braket}
\usepackage{graphicx}
\usepackage{iopams}
\usepackage{eufrak}
\usepackage{float}
\DeclareMathAlphabet{\mathpzc}{OT1}{pzc}{m}{it}
\begin{document}

\title[Auto- and Cross-Correlation-Functions as entanglement quantifiers in semiconductor microcavities.]
{Auto- and Cross-Correlation-Functions as entanglement quantifiers in semiconductor microcavities.}

\author{Rub\'{e}n.~Dar\'{i}o.~Guerrero-Mancilla}

\address{\centerline{\emph{Grupo de \'Optica e Informaci\'on Cu\'antica,
Departamento de F\'{\i}sica,}} \centerline{\emph{Universidad
Nacional de Colombia - Bogot\'a }}} \ead{rdguerrerom@unal.edu.co}

\begin{abstract}
\noindent{\footnotesize .}
                          The dynamics of the exciton-photon entanglement, in
a semiconductor microcavity is analyzed. Finding a closed analytical expression
for the time evolution of the concurrence. Using as model two coupled, quantum
oscillators with detuning between them. Supposing that the system is at a
temperature closer to zero. And that the quantum state of the system, remains in
the one excitation sector of the total Hilbert space. The former theoretical setup,
describes the excitons coupled to the photons in the microcavity accurately, for
a system with ultra low density of excitons. We remark in first place, that our
closed expression for the concurrence dynamics. Enable us to establish a good
analytical criterion, to determine the coupling dynamical regime, in which the
system is. Even more, we show that our criterion, is in good agreement with
typically accepted, experimental signatures for strong, and wake coupling regimes.
Finally we present theoretical evidence, suggesting that the entanglement between
the excitons and the photons. Can be measured in experiments, via the cross
correlation contribution, to the time resolved spectra of the system.
\end{abstract}


\maketitle
\section{Introduction} 
Quantum information science \cite{Nielsen}, study the information processing tasks that can be accomplished using quantum mechanical systems. And the entanglement, is a uniquely quantum mechanical resource, that play a key role, in roughly all of the most interesting applications of quantum information science. One of the quantum system configurations, where is best know how to quantify the amount of entanglement, between two component subsystems, is a two qubits system. Where the amount of this physical resource, is measured by the Wotters concurrence \cite{Wootters1998a}. However to determine the concurrence experimentally. Total or partial access, to the  quantum state of the  system is quintessential. Recently various proposals had made, to achieve this challenge, in a two qubits system. Quantum tomography \cite{Zoller1997} is one of the most popular approaches, because was the first one, successfully implemented for light polarization degrees of freedom \cite{White1999}. But have some potentially disadvantaging features, one of them is that have  many copies of the  state of the system at hand is mandatory. In order to improve the reliability  of the results. Other disadvantage is that the  information of system is retrieved, through local projective measurements in a quantum states basis. That for the most general situations could be ambiguously specified \cite{Bagan2005}. Therefore, great effort had made in order of improve the quantum tomography technique. Among them, the Ancilla-Assisted Quantum Process Tomography \cite{Altepeter}. That proposal, reduces dramatically, the mandatory number of copies of the state of the system, that must be available, to obtain reliable results. At expenses of the coupling to a third party called ancilla. The direct characterization of quantum dynamics \cite{Mohseni2006}. That relies on quantum error correction algorithms,  and still  having as a very hard feature. The unavoidable need, of have at hand a big number of copies, of  the state of the system. In order to produce reliable information about it. All the former approaches has been designed to, and tested in, full optical quantum systems. Despite of that,  some proposals were making for  quantum states in other kind of architectures too, by example for solid state systems \cite{Samuelsson2006},  nonlinear optical systems \cite{Altepeter2005}, and for cavity QED systems \cite{Lee2008}. However nowadays, one of the most pragmatic and promising goals of the quantum information field, is build up interfaces between high speed flying qubits, as the photons used for quantum channel communications,  and a stationary solid state qubits system. Performing quantum information processing tasks \cite{Gisin2010}. Material excitations in solid state systems, as the exciton for example, are promising candidates for solid states qubits. Because, not only offer very good coupling with the light field in a semiconductor microcavity --reason for which we call this component in our model, the matter emitter--. But also,  because the exciton constitute a  solid state system, that could provide on chip integration with other quantum computation hardware.
The principal aim of this paper, is study the dynamics of the matter-field entanglement in a semiconductor microcavity at a very low temperature. By means the use of a ``Simplified'' model, of two coupled and detuned  oscillators. That we use to provide insight to, an experimentally observable, and single copy based, measurement of the matter-field entanglement. V\'ia the cross correlation spectra of the system. With this goal in perspective, we organize the paper as follow.  In the second section we describe the theoretical model, its principal features, limitations and sketch, how the evolution in time of the state of the system, and the emission spectra was obtained. In the third section we  present the principal results of the paper.  Deriving a closed expression for concurrence, and showing why the cross correlation function, carries information about the exciton-light entanglement in the microcavity. That other spectrals measurements like the photoluminescence spectra, cannot provide. In the fourth section we explain some features, and implications of our result for the concurrence. Deriving from it, in first place an  analytical criterion based on the experimentally controllable parameters of the system, to be in the strong, or weak  coupling regime of the dynamics. And using the photoluminescence, we show agreement of our criterion with often accepted signatures of the strong coupling regime. In second place we show the times of maximum and minimum concurrence as function of the experimentally controllable parameters, obtained from our closed expression for the concurrence. In the fifth section some concluding remarks was  done.
 \section{Theoretical model and  dynamics of the system}
 With the aim of generalize the description, to other kind of bosonic systems, we will refer by matter emitter, to the excitations in the material. Excitons in the ultralow density limit are a particular case of which. We model the  system photons-excitons introducing a harmonic oscillator of frequency $\Omega$ described by bosonic operators ${\hat e}$ $({\hat e}^{\dagger})$ for the destruction (creation) of an excitation in the matter emitter, and coupling the former oscillator to a second one of frequency $\omega$ described by bosonic operators ${\hat a}$ $({\hat a}^{\dagger})$ for the destruction (creation) of a photon in the microcavity. Our interest here is study dynamical processes in which the energy is approximately conserved, thus the dynamics of   the coupling between the two oscillators was taken in such way that rapidly rotating terms banish in the so called rotating wave approximation, we take the emitter-photons interaction proportional to a coupling parameter $g$ that measure the rate of energy exchange between the oscillators, with  the former considerations, the  Hamiltonian of the system can be written as
\begin{equation}\label{eq:Ham_Free}
\hat{H} = \hbar\Omega{\hat e}^{\dagger}{\hat e} +\hbar\omega{\hat
a}^{\dagger}{\hat a} + \hbar g ({\hat a}^{\dagger}{\hat e}+{\hat a}{\hat e}^{\dagger}),
\end{equation}
where the frequency difference $\Delta=\Omega -\omega $ is often called the detuning parameter. With the aim of include effects of finite linewidth for the excitations in the system, we introduce the parameter  $\kappa$, that is inversely proportional to the lifetime of an excitation in the matter emitter. And a parameter $\gamma$, that is inversely proportional to the lifetime of an excitation of the field in the microcavity. Therefore, we describe the  dynamics of the state of  the system,  using a phenomenological Liouvillian in the Lindblad form, at zero temperature and in the Markov approximation.  Where each oscillator is initially in a coherent state
\begin{equation}
\label{eq:lvneq} \frac{d{\hat \rho}}{dt}= -\frac{i}{\hbar}[{\hat
H},{\hat \rho}] + \gamma (2{\hat a}{\hat \rho}{\hat a}^{\dagger}
-{\hat a}^{\dagger}{\hat a}{\hat \rho} -{\hat \rho}{\hat
a}^{\dagger}{\hat a})+ \kappa (2{\hat e}{\hat \rho}{\hat
e}^{\dagger} -{\hat e}^{\dagger}{\hat e}{\hat \rho} -{\hat
\rho}{\hat e}^{\dagger}{\hat e}).
\end{equation}
This formulation of the dynamics have desirable properties well established in the literature as hermiticity, trace conservation and have semipositive eigenvalues for the  state of the system \cite{Lindblad1976a}. To find the system's dynamics in (\ref{eq:lvneq}) we use the method proposed in \cite{mancilla08}  postulating the \emph{Anzats}
\begin{equation}
\label{ansatz}
 \hat{\rho}(t) = {\mathcal D}(\beta(t),\alpha(t))
\tilde{\rho}(t) {\mathcal D}^\dag(\beta(t),\alpha(t)),
\end{equation}
where  ${\mathcal D}(\beta(t),\alpha(t))$ is an field-matter displacement operator acting in the two modes with complex coherent amplitude  $\alpha(t)$ and $\beta(t)$ for the field and the matter emitter respectively, and  $\tilde{\rho}(t)$ is a state of the full system in the interaction picture defined by the field-matter displacement operator which time derivative was reported in \cite{mancilla08}; the \emph{Anzats}  in the equation (\ref{ansatz}) enable us to decouple the evolution of the coherent amplitude, from the evolution of the  state of the system, finding the following system of differentials equations for the evolution of  the  state of the system, the respective equations for the coherent amplitude is reported in equation (\ref{eq:dDisplacementdt}) in  the appendix 2
\begin{eqnarray} \label{eq:dRhotilde}
\frac{d{\hat {\tilde \rho}}}{dt}= -\frac{i}{\hbar}[{\hat
H},{\hat {\tilde \rho}}] + \gamma (2{\hat a}{\hat \rho}{\hat a}^{\dagger}
-{\hat a}^{\dagger}{\hat a}{\hat {\tilde \rho}} -{\hat {\tilde \rho}}{\hat
a}^{\dagger}{\hat a})+ \kappa (2{\hat e}{\hat {\tilde \rho}}{\hat
e}^{\dagger} -{\hat e}^{\dagger}{\hat e}{\hat {\tilde \rho}} -{\hat
\rho}{\hat e}^{\dagger}{\hat e}).
\end{eqnarray}
Quantum effects like entanglement and purity are completely encoded in the  state of the system ${\hat {\tilde \rho}}$. We have choosen as initial state of the field in the microcavity, a coherent state $\hat{\rho}_2(t) = \textrm{tr}_1\hat{\rho}(t)=\Ket{\alpha_0}\Bra{\alpha_0}$ choosing the amplitude in such way that $\mid \alpha_0 \mid \leq 1$ that in the  case of incoherent pumping is not present is not a demanded restriction. In the other hand, the matter emitter is initially in a pure state that is linear combination of the states with zero and one excitation it is
\begin{equation} \label{ec:initialystate1}\fl
\label{eq:InitialState} {\mathcal D}\left(0,\alpha_0 \right)
\underbrace{ \left(\cos(\theta)\ket{0}+\sin(\theta)\ket{1}\right)
\left(\cos(\theta)\bra{0}+\sin(\theta)\bra{1}\right) \otimes
\ket{0}\bra{0}}_{\tilde{\rho}(0)} {\mathcal D}^\dag\left(0,\alpha_0
\right),
\end{equation}
if the initial state of the matter emitter is the excitations vacuum, the system stair in a state that is  pure, separable and equal to $\rho(t)=\ket{\beta(t)}\bra{\beta(t)}\otimes\ket{\alpha(t)}\bra{\alpha(t)}.$  Writing the  state of the system in the base of zero and one excitation in each subsystem, it can be ordered as the density matrix for  two qubits
\begin{equation}\label{eq:Rhotilde}
\tilde{\rho}(t) = \sum_{i_1 i_2 j_1 j_2} \tilde{\rho}_{i_2 i_1}^{j_2
j_1} \ket{i_1 i_2} \bra{j_1 j_2}=\left(
\begin{array}{llll}
 \tilde{\rho}_{00}^{00}(t) & \tilde{\rho}_{00}^{01}(t) & \tilde{\rho}_{00}^{10}(t) & 0 \\
 \tilde{\rho}_{01}^{00}(t) & \tilde{\rho}_{01}^{01}(t) & \tilde{\rho}_{01}^{10}(t) & 0 \\
 \tilde{\rho}_{10}^{00}(t) & \tilde{\rho}_{10}^{01}(t) & \tilde{\rho}_{10}^{10}(t) & 0 \\
 0 & 0 & 0 & 0
\end{array}
\right).
\end{equation}
Where we truncate the Hilbert space to one excitation sector. It is we assume that elements of the density matrix, with more than one excitation are zero. The former assumption remains valid for all times in the evolution, for  initial conditios with $\mid \alpha_0 \mid \ll 1$, $\mid \beta_0 \mid \ll 1$, and the total number of excitations in the excitons-photons initial state equal to one almost. For this set of initial conditions, the system state dynamics occurs fully in the one excitation sector of the Hilber space of the system. Because the sectors of the density matrix with more than one excitation does not populate significantly. With the aim of include finite linewidth in the  time evolution of the  state of the system. We introduce the Liouville super-operator \cite{Lindblad1976a}
\begin{equation}
\label{eq:lvneqip} {\hat{\hat {\mathcal L}}}(\bullet) =
-\frac{i}{\hbar}[{\hat H},\bullet] + \gamma (2{\hat
a}\bullet{\hat a}^{\dagger} -\{ {\hat a}^{\dagger}{\hat
a},\bullet\})+ \kappa (2{\hat e}\bullet{\hat e}^{\dagger} -\{ {\hat
e}^{\dagger}{\hat e},\bullet\}).
\end{equation}
In such way that we must confront the differential equation $\dot{\tilde{\rho}}(t)={\hat{\hat {\mathcal L}}}(\tilde{\rho}(t))$. To solve it, we appeal to the Laplace transform properties. Thus taking the Laplace transform to both sides  of the differential equation --here represented as $ \widehat{\tilde{\rho}}(s)$-- we have ${s \widehat{\tilde{\rho}}}(s)+{\hat{\hat {\mathcal
L}}}({\widehat{\tilde{\rho}}}(s))=\tilde{\rho}(0).$ That live in our hands,  the simpler problem of solve a system of linear equations  $(s\mathbb{I} +{\mathfrak L})
\overrightarrow{{\widehat{\tilde{\rho}}}(s)}=\overrightarrow{\tilde{\rho}(0)}$ where the  state of the system has been reordered by his rows  as a larger vector and the Liouville super-operator as the matrix
\begin{equation} \fl {\tiny{\mathfrak L}=\left(
\begin{array}{ccccccccc}
 0 & 0 & 0 & 0 & -2 \kappa  & 0 & 0 & 0 & -2 \gamma  \\
 0 & \kappa +i \Omega  & -i g & 0 & 0 & 0 & 0 & 0 & 0 \\
 0 & -i g & \gamma +i \omega  & 0 & 0 & 0 & 0 & 0 & 0 \\
 0 & 0 & 0 & \kappa -i \Omega  & 0 & 0 & i g & 0 & 0 \\
 0 & 0 & 0 & 0 & 2 \kappa  & -i g & 0 & i g & 0 \\
 0 & 0 & 0 & 0 & -i g & \gamma +\kappa +i (\omega -\Omega ) & 0 & 0 & i
   g \\
 0 & 0 & 0 & i g & 0 & 0 & \gamma -i \omega  & 0 & 0 \\
 0 & 0 & 0 & 0 & i g & 0 & 0 & \gamma +\kappa -i (\omega -\Omega ) & -i
   g \\
 0 & 0 & 0 & 0 & 0 & i g & 0 & -i g & 2 \gamma
\end{array}
\right),}
\end{equation}
the subsidiary condition  $Det((s\mathbb{I} +{\mathfrak L}))\neq 0$ for the existence of the solution is fulfilled, thus the evolution of the state of the system can be find as the inverse Laplace transform of $\overrightarrow{{\widehat{\tilde{\rho}}}(s)}=(s\mathbb{I}
+{\mathfrak L})^{-1}\overrightarrow{\tilde{\rho}(0)}$. The former is the simpler particular situation of the case in which the systems is affected by coherent pumping of photons, in this situation we take the Hamiltonian of the system as
\begin{equation}
\fl\hat{H}_{_\mathcal{P}} = \hbar\Omega{\hat e}^{\dagger}{\hat e} +\hbar\omega{\hat
a}^{\dagger}{\hat a} + \hbar g ({\hat a}^{\dagger}{\hat e}+{\hat a}{\hat e}^{\dagger})+\hbar \mathcal{F}\left({\hat a} e^{i \omega_{_D}t}+{\hat a}^{\dagger} e^{-i \omega_{_D}t}\right),
\end{equation}
where the parameter $\mathcal{F}$ represents the photons pumping intensity. Switch  into the interaction picture defined by the auxiliary Hamiltonian ${\hat \mathcal{H}}_0=\hbar \omega_{_D}\left({\hat e}^{\dagger}{\hat e}+{\hat a}^{\dagger}{\hat a}\right)$ enable us to write the  Hamiltonian of the system for this case without time dependence
\begin{equation}
\fl\hat{\tilde H}_{_\mathcal{P}} = \hbar\delta{\hat e}^{\dagger}{\hat e} +\hbar\left(\delta+\Delta\right){\hat
a}^{\dagger}{\hat a} + \hbar g ({\hat a}^{\dagger}{\hat e}+{\hat a}{\hat e}^{\dagger})+\hbar \mathcal{F}\left({\hat a} +{\hat a}^{\dagger}\right),
\end{equation}
where a new parameter for the detuning respect to the pumping frequency $\delta=\omega-\omega_{_D}$ has been introduced, and the matter-field detuning parameter $\Delta$ have the usual meaning. With the aim of include finite linewidth effects, we again describe the time evolution of the  state of the system by means the following zero temperature Liouville super-operator in the Markov approximation
\begin{equation}
\label{eq:lvneqip_cp} \fl {\hat{\hat {\mathcal L}}}_{_\mathcal{P}}(\bullet) =
-\frac{i}{\hbar}[\hat{H}_{_\mathcal{P}},\bullet] + \gamma (2{\hat
a}\bullet{\hat a}^{\dagger} -\{ {\hat a}^{\dagger}{\hat
a},\bullet\})+ \kappa (2{\hat e}\bullet{\hat e}^{\dagger} -\{ {\hat
e}^{\dagger}{\hat e},\bullet\}),
\end{equation}
thus  again the use of the \emph{Anzats} in the equation (\ref{ansatz}) enable us decouple the time evolution for the coherent amplitude and the  time evolution of the state of the system with the pumping affecting just the coherent amplitudes, in this case the time evolution of the state is
\begin{eqnarray}
\label{eq:dRhotilde_pump}\fl \frac{d}{dt}{\hat {\tilde \rho}}={\hat{\hat {\mathcal L}}}_{_\mathcal{PR}}({\hat {\tilde \rho}})
\end{eqnarray}
the coherent amplitude evolves according the equation (\ref{eq:dDisplacementdt_pump}) in the appendices, that must be solved using the initial conditions in the equation (\ref{ec:initialystate1}) taking again $\alpha(0)=\alpha_0$ and $\beta(0)=0$. The Liouvillian that we find for evolution of the system can be written as
\begin{equation}
\label{eq:lvneqip_cpr} \fl {\hat{\hat {\mathcal L}}}(\bullet) =
-\frac{i}{\hbar}[\hat{H}_{_\mathcal{PR}},\bullet] + \gamma (2{\hat
a}\bullet{\hat a}^{\dagger} -\{ {\hat a}^{\dagger}{\hat
a},\bullet\})+ \kappa (2{\hat e}\bullet{\hat e}^{\dagger} -\{ {\hat
e}^{\dagger}{\hat e},\bullet\}),
\end{equation}
where an  effective Hamiltonian has been defined in the following way
 \begin{equation}
\fl\hat{\tilde H}_{_\mathcal{PR}} = \hbar\delta{\hat e}^{\dagger}{\hat e} +\hbar\left(\delta+\Delta\right){\hat
a}^{\dagger}{\hat a} + \hbar g ({\hat a}^{\dagger}{\hat e}+{\hat a}{\hat e}^{\dagger}).
\end{equation}
Finally we remark that, the replacement rules $\Omega\rightarrow \delta $ and $\omega\rightarrow \delta +\Delta$ relate the equation (\ref{eq:dRhotilde}) to the equation (\ref{eq:dRhotilde_pump}) which solution has been found before in the case without coherent pumping.
\subsection{First order coherence functions}
We define as a set of  first order coherence functions, a  set of relevant two times operators of the form
\begin{equation}
\fl G^{(1)}_{\hat O}(t,\tau)=\langle {\hat O}^{\dagger}(t) {\hat O}(t+\tau)\rangle,
\end{equation}
where  ${\hat O}$ represents operators acting on the first or second harmonic oscillators, in our particular case, result of relevance the following set of coherence functions
\begin{eqnarray}
\fl  G^{(1)}_{_{C}}(t,\tau)=\langle {\hat a}^{\dagger}(t) {\hat a}(t+\tau)\rangle\\
\fl  G^{(1)}_{_{X}}(t,\tau)=\langle {\hat e}^{\dagger}(t) {\hat e}(t+\tau)\rangle \\
\fl  G^{(1)}_{_{Cr_1}}(t,\tau)=\langle {\hat a}^{\dagger}(t) {\hat e}(t+\tau)\rangle \\
\fl  G^{(1)}_{_{Cr_2}}(t,\tau)=\langle {\hat e}^{\dagger}(t) {\hat a}(t+\tau)\rangle
\end{eqnarray}
where  $G^{(1)}_{_{C}}(t,\tau)$ measure the contribution to the  photoluminescence spectra, that is result of stimulated emission processes in the matter emitter,  $G^{(1)}_{_{X}}(t,\tau)$ measure  the contribution to the photoluminescence spectra,  that is result of spontaneous emission processes  in the matter emitter, and finally  $G^{(1)}_{_{Cr}}(t,\tau)$ measure the cross correlations spectra, originated by the energy feedback between the filed in the microcavity, and the matter emitter. To obtain this coherence functions we appeal to the quantum regression theorem,  the theorem states that if  the evolution of the single operator $\langle{\hat O_i(\tau)}\rangle$ is determined by the following equation
\begin{eqnarray}
\fl  \frac{d}{d\tau}\langle{\hat O_i(\tau)}\rangle=\sum_jC_{j,i}\langle{\hat O_j(\tau)}\rangle
\end{eqnarray}
where $C_{j,i}$ are a set of fixed coefficients determined by the  dynamics of the system, thus the same set of coefficients enable us to determine the dynamics of the coherence functions
\begin{eqnarray}
\fl  \frac{d}{d\tau}\langle{\hat O_k(t)}^{\dagger}{\hat O_i(t+\tau)}\rangle=\sum_j C_{j,i}\langle{\hat O_k(t)}^{\dagger}{\hat O_j(t+\tau)}\rangle.
\end{eqnarray}
In the present situation the coefficients $C_{j,i}$ are determined in the Heisenberg picture by the Liouville super-operator 
\begin{equation}
\label{eq:lvneqip} \fl  {\hat{\hat {\mathcal L}}}(\bullet) =
\frac{i}{\hbar}[{\hat H},\bullet] + \gamma (2{\hat
a}\bullet{\hat a}^{\dagger} -\{ {\hat a}^{\dagger}{\hat
a},\bullet\})+ \kappa (2{\hat e}\bullet{\hat e}^{\dagger} -\{ {\hat
e}^{\dagger}{\hat e},\bullet\}),
\end{equation}
thus representing the former Liouvillian as a super-operator in the two qubits basis, truncated to the one excitation sector of the Hilbert space, by the matrix ${\mathfrak L},$ this matrix representation of the Liouvillian can be written in the following way
{\small \begin{equation}
\fl {\mathfrak L}=\left(
\begin{array}{ccccccccc}
 0 & 0 & 0 & 0 & -2 \kappa  & 0 & 0 & 0 & -2 \gamma  \\
 0 & \kappa +i \Omega  & -i g & 0 & 0 & 0 & 0 & 0 & 0 \\
 0 & -i g & \gamma +i \omega  & 0 & 0 & 0 & 0 & 0 & 0 \\
 0 & 0 & 0 & \kappa -i \Omega  & 0 & 0 & i g & 0 & 0 \\
 0 & 0 & 0 & 0 & 2 \kappa  & -i g & 0 & i g & 0 \\
 0 & 0 & 0 & 0 & -i g & \gamma +\kappa +i (\omega -\Omega ) & 0 & 0 & i
   g \\
 0 & 0 & 0 & i g & 0 & 0 & \gamma -i \omega  & 0 & 0 \\
 0 & 0 & 0 & 0 & i g & 0 & 0 & \gamma +\kappa -i (\omega -\Omega ) & -i
   g \\
 0 & 0 & 0 & 0 & 0 & i g & 0 & -i g & 2 \gamma
\end{array}
\right).
\end{equation}}
The operators on which ${\mathfrak L}$ can act are represented by column vectors, as an example ${\hat a}\rightarrow \overrightarrow{{\mathfrak a}}=(0,0,1,0,0,0,0,0,0)^{\dagger}$, ${\hat a}^{\dagger}\rightarrow \overrightarrow{{\mathfrak a}}^{\star}=(0,0,0,0,0,0,1,0,0)^{\dagger}$, ${\hat e}\rightarrow \overrightarrow{{\mathfrak e}}=(0,1,0,0,0,0,0,0,0)^{\dagger}$, and ${\hat e}^{\dagger}\rightarrow \overrightarrow{{\mathfrak e}}^{\star}=(0,0,0,1,0,0,0,0,0)^{\dagger}$. The former super-operator determines the time evolution of the  relevant operators of the system, enabling us to obtain the coefficients involved in the quantum regression theorem
\begin{eqnarray}
\fl \frac{d}{d\tau}\langle{\hat a}\rangle=\left(\overrightarrow{\mathfrak{a}}\cdot{\mathfrak L}\cdot\overrightarrow{\mathfrak{a}}\right)\langle{\hat a(\tau)}\rangle+\left(\overrightarrow{\mathfrak{e}}\cdot{\mathfrak L}\cdot\overrightarrow{\mathfrak{a}}\right)\langle{\hat e(\tau)}\rangle=(\gamma +i \omega) \langle{\hat a(\tau)}\rangle-i g \langle{\hat e(\tau)}\rangle\\
\fl \frac{d}{d\tau}\langle{\hat e}\rangle=\left(\overrightarrow{\mathfrak{e}}\cdot{\mathfrak L}\cdot\overrightarrow{\mathfrak{e}}\right)\langle{\hat e(\tau)}\rangle+\left(\overrightarrow{\mathfrak{a}}\cdot{\mathfrak L}\cdot\overrightarrow{\mathfrak{e}}\right)\langle{\hat e(\tau)}\rangle=(\kappa +i \Omega)\langle{\hat e(\tau)}\rangle-i g \langle{\hat a(\tau)}\rangle,
\end{eqnarray}
in such way that we find straightforwardly the following set of differential equations for the first order coherence functions
 \begin{equation}
\left(
   \begin{array}{c}
     \dot{G}^{(1)}_{_{C}}(t,\tau) \\
     \dot{G}^{(1)}_{_{X}}(t,\tau) \\
     \dot{G}^{(1)}_{_{Cr_1}}(t,\tau) \\
     \dot{G}^{(1)}_{_{Cr_2}}(t,\tau)^{\star} \\
   \end{array}
 \right)=\left(
\begin{array}{cccc}
 \gamma +i \omega  & 0 & -i g & 0 \\
 0 & \kappa +i \Omega  & 0 & -i g \\
 -i g & 0 & \kappa +i \Omega  & 0 \\
 0 & -i g & 0 & \gamma +i \omega
\end{array}
\right)\left(
           \begin{array}{c}
     G^{(1)}_{_{C}}(t,\tau) \\
     G^{(1)}_{_{X}}(t,\tau) \\
     G^{(1)}_{_{Cr_1}}(t,\tau) \\
     G^{(1)}_{_{Cr_2}}(t,\tau) \\
           \end{array}
         \right).
 \end{equation}
Defining the coefficients matrix as $\mathfrak{A}$, we appeal to the Laplace Transform method to look for the only one solution, in such way that the coherence functions will be determined by the inverse Laplace transform of the matrix $(\mathfrak{A}+s\mathbb{I})^{-1}$ acting on the set of initial conditions
\begin{equation}
\left(
           \begin{array}{c}
     G^{(1)}_{_{C}}(t,0) \\
     G^{(1)}_{_{X}}(t,0) \\
     G^{(1)}_{_{Cr_1}}(t,0) \\
     G^{(1)}_{_{Cr_2}}(t,0) \\
           \end{array}
         \right)=\left(
           \begin{array}{c}
     \langle {\hat a}^{\dagger}(t) {\hat a}(t)\rangle \\
     \langle {\hat e}^{\dagger}(t) {\hat e}(t)\rangle\\
     \langle {\hat a}^{\dagger}(t) {\hat e}(t)\rangle \\
     \langle {\hat e}^{\dagger}(t) {\hat a}(t)\rangle\\
           \end{array}
         \right)=\left(
           \begin{array}{c}
     \langle {\hat a}^{\dagger} {\hat a}(t)\rangle \\
     \langle {\hat e}^{\dagger} {\hat e}(t)\rangle\\
     \langle {\hat a}^{\dagger} {\hat e}(t)\rangle \\
     \langle {\hat e}^{\dagger} {\hat a}(t)\rangle\\
           \end{array}
         \right)=\left(
           \begin{array}{c}
     \rho^{10}_{10}(t)\\
     \rho^{01}_{01}(t)\\
     \rho^{01}_{10}(t)\\
     \rho^{10}_{01}(t)\\
           \end{array}
         \right)
\end{equation}
where in the intermediate step, the properties of the dynamical semigroup has been used \cite{Lindblad1976a}.  Defining the constant ${\tilde G}=\sqrt{-4 g^2+(\gamma -\kappa +i (\omega -\Omega ))^2}$ we find the following solutions, for the set of coherence functions
{\tiny \begin{eqnarray}\label{eq:correlations}
\fl G^{(1)}_{_{C}}(t,\tau)&= \frac{e^{-\frac{1}{2} \tau  (\gamma +\kappa +i (\omega +\Omega ))}{\tilde G}^{\star} \left(\rho^{10}_{10}(t) {\tilde G} \cosh \left(\frac{{\tilde G} \tau}{2}\right)+\sinh \left(\frac{{\tilde G} \tau }{2}\right) (\rho^{10}_{10}(t) (-\gamma +\kappa -i \omega +i \Omega )+2 i \rho^{01}_{10}(t) g)\right)}{{\tilde G}^{\star}{\tilde G}} \\
\fl G^{(1)}_{_{X}}(t,\tau)&=\frac{e^{-\frac{1}{2} \tau  (\gamma +\kappa +i (\omega +\Omega ))} {\tilde G}^{\star} \left(\rho^{01}_{01}(t) {\tilde G} \cosh \left(\frac{{\tilde G}\tau}{2}\right)+\sinh \left(\frac{{\tilde G} \tau }{2}\right) (2 i \rho^{10}_{01}(t) g+\rho^{01}_{01}(t) (\gamma -\kappa +i (\omega -\Omega)))\right)}{{\tilde G}^{\star}{\tilde G}}\\
\fl G^{(1)}_{_{Cr_1}}(t,\tau)&=\frac{e^{-\frac{1}{2} \tau  (\gamma +\kappa +i (\omega +\Omega ))}{\tilde G}^{\star}\left(\rho^{01}_{10}(t) {\tilde G} \cosh \left(\frac{{\tilde G} \tau}{2}\right)+\sinh \left(\frac{{\tilde G}\tau }{2}\right) (2 i \rho^{10}_{10}(t) g+\rho^{01}_{10}(t) (\gamma -\kappa +i (\omega -\Omega)))\right)}{{\tilde G}^{\star}{\tilde G}}\\
\fl G^{(1)}_{_{Cr_2}}(t,\tau)&=\frac{e^{-\frac{1}{2} \tau  (\gamma +\kappa +i (\omega +\Omega ))}{\tilde G}^{\star} \left(\rho^{10}_{01}(t) {\tilde G} \cosh \left(\frac{{\tilde G} \tau}{2}\right)+\sinh \left(\frac{{\tilde G} \tau }{2}\right) (\rho^{10}_{01}(t) (-\gamma +\kappa -i \omega +i \Omega )+2 i \rho^{01}_{01}(t) g)\right)}{{\tilde G}^{\star}{\tilde G}}.
\end{eqnarray}}
According to the Wiener-Khintchine theorem, the contribution to the spectra of any of the coherence functions can be obtained as
\begin{equation}\label{eq:Kinchin}
I_{l}(t,\omega_{_{F}})=\frac{1}{\pi} \mathfrak{Re}\left\{\int_{0}^{\infty}e^{i\omega_{_{F}} \tau }G^{(1)}_{l}(t,\tau)d\tau\right\}.
\end{equation}
with $l$ taking values $\{C,X,Cr_1,Cr_2\}$, in our over-simplified model of coupled  oscillators, the former integral can be evaluated  analytically  for each one of the system's coherence functions.
\clearpage
\section{Results}
\begin{figure}[h!b!]
  \begin{center}
  \includegraphics[width=10 cm]{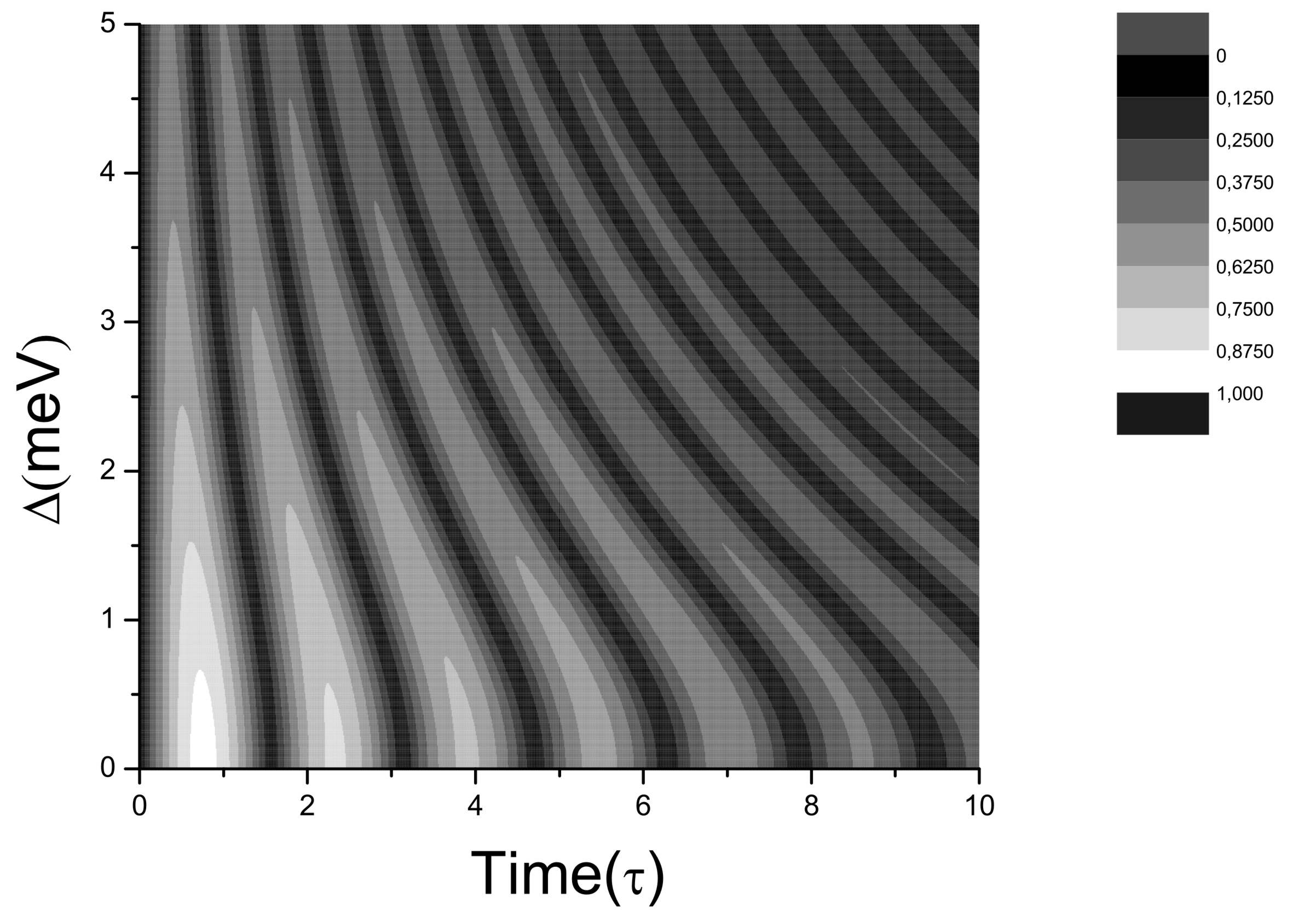}\\
  \caption{Concurrence as function of time and detuning in the strong coupling regime (\ref{eq:free_concurrence}), with $g= 1$ meV, $\gamma=0.1$ meV and $\theta=\pi/2;$ we choose for this graphic the parameter $\kappa=0.01$ meV  that fulfill the conditions established in the equation (\ref{Cond_SC}). }\label{ODR}
  \end{center}
\end{figure}

\begin{figure}[h!t!]
\begin{center}$
\begin{array}{cc}
\includegraphics[width=6.75cm]{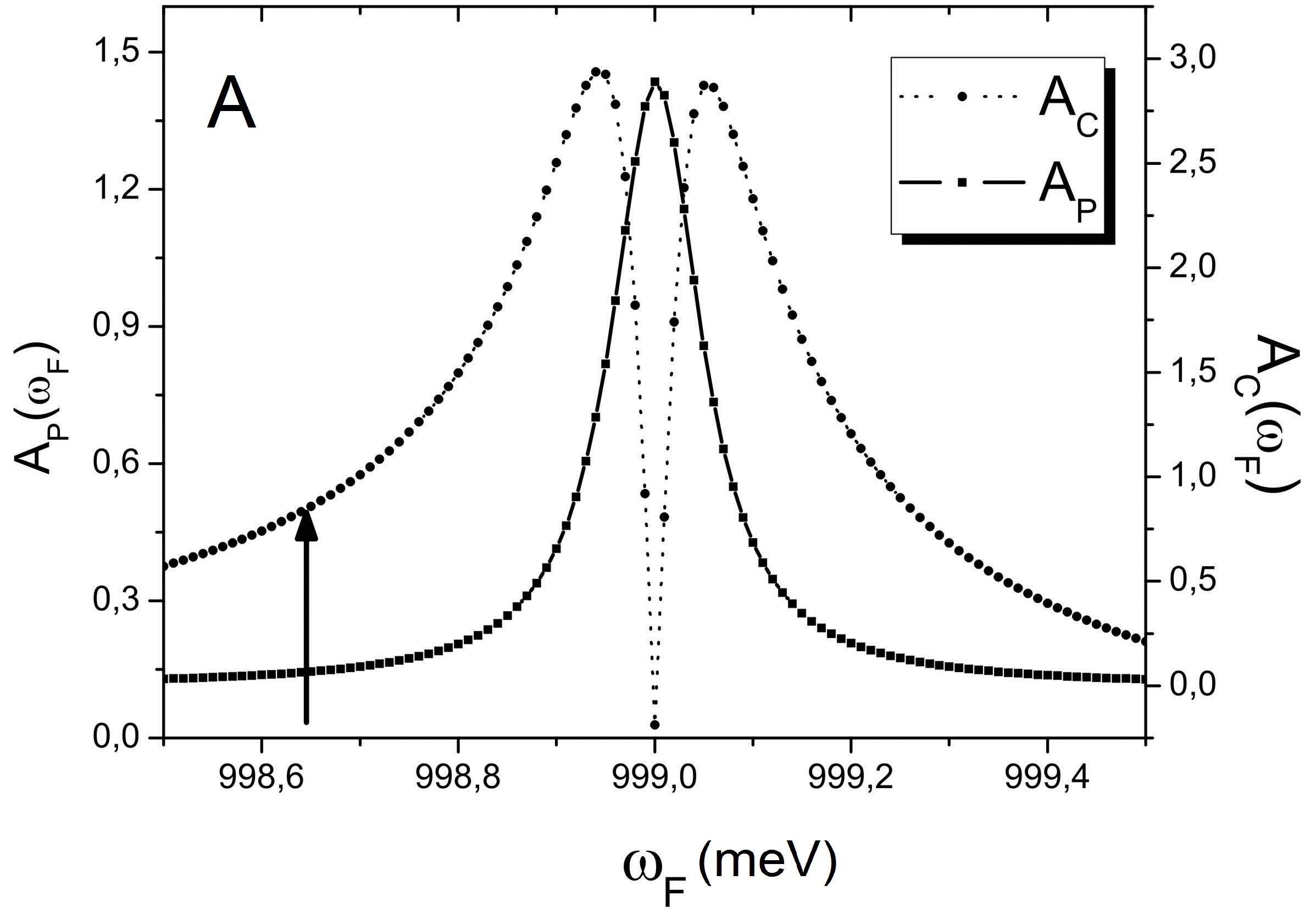} &
\includegraphics[width=6.75cm]{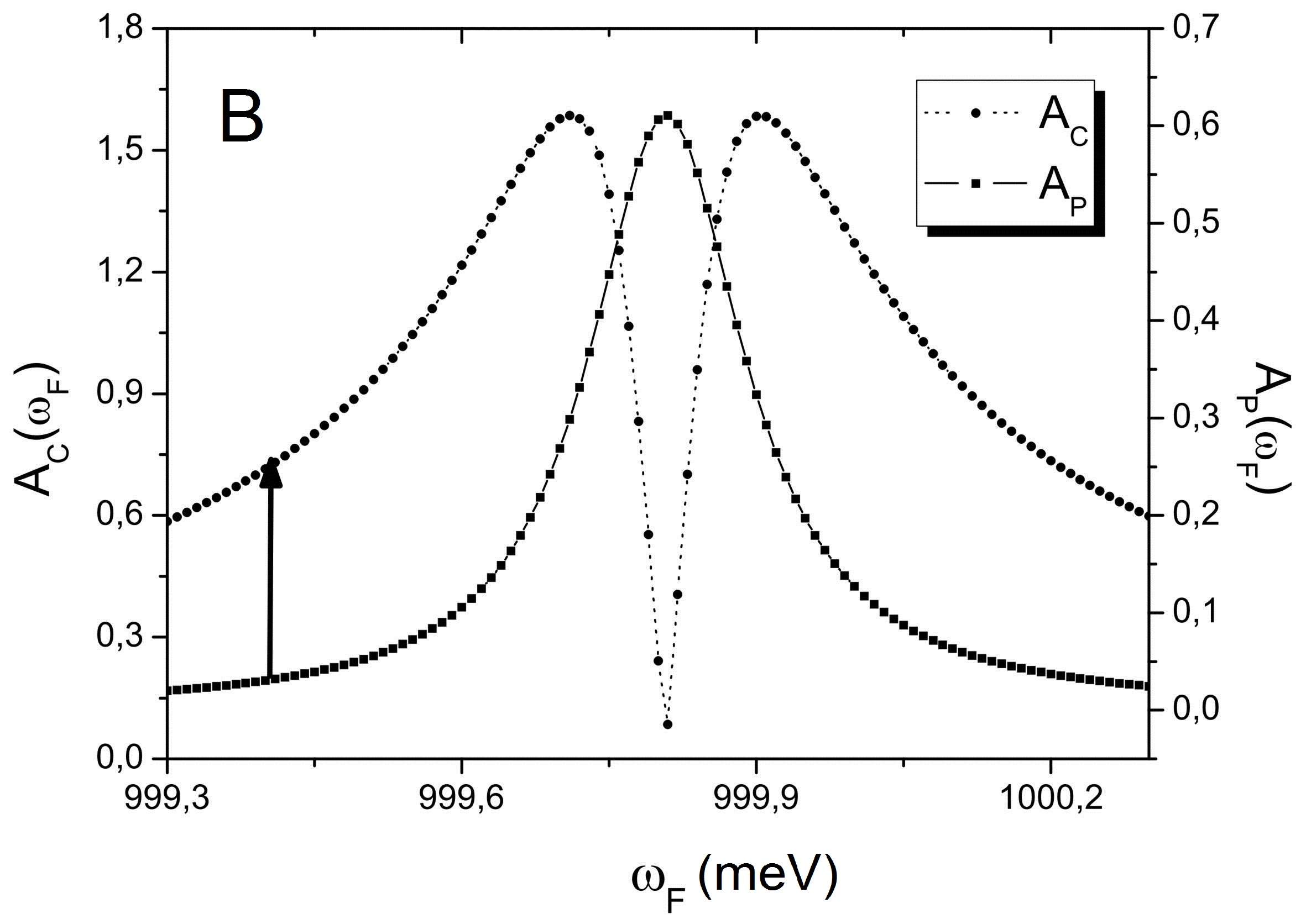}
\end{array}$
\caption{Absolute value of the amplitude $A_C$ and $A_P$ at resonance $(\Delta=0)$ in A and with detuning $(\Delta=5$ meV$)$ in B. Using the parameter values $g= 1 $ meV,  $\theta=\pi/2$, $\omega=1 $ eV, $\kappa=0.01 $ meV, and $\gamma=0.1 $ meV. In the zone signaled by the arrows, the contribution of the coherence is always grater than the population contribution.}\label{Expl_criteria}
\end{center}
\end{figure}

\begin{figure}[h!b!]
  \begin{center}
  \includegraphics[width=12.5 cm]{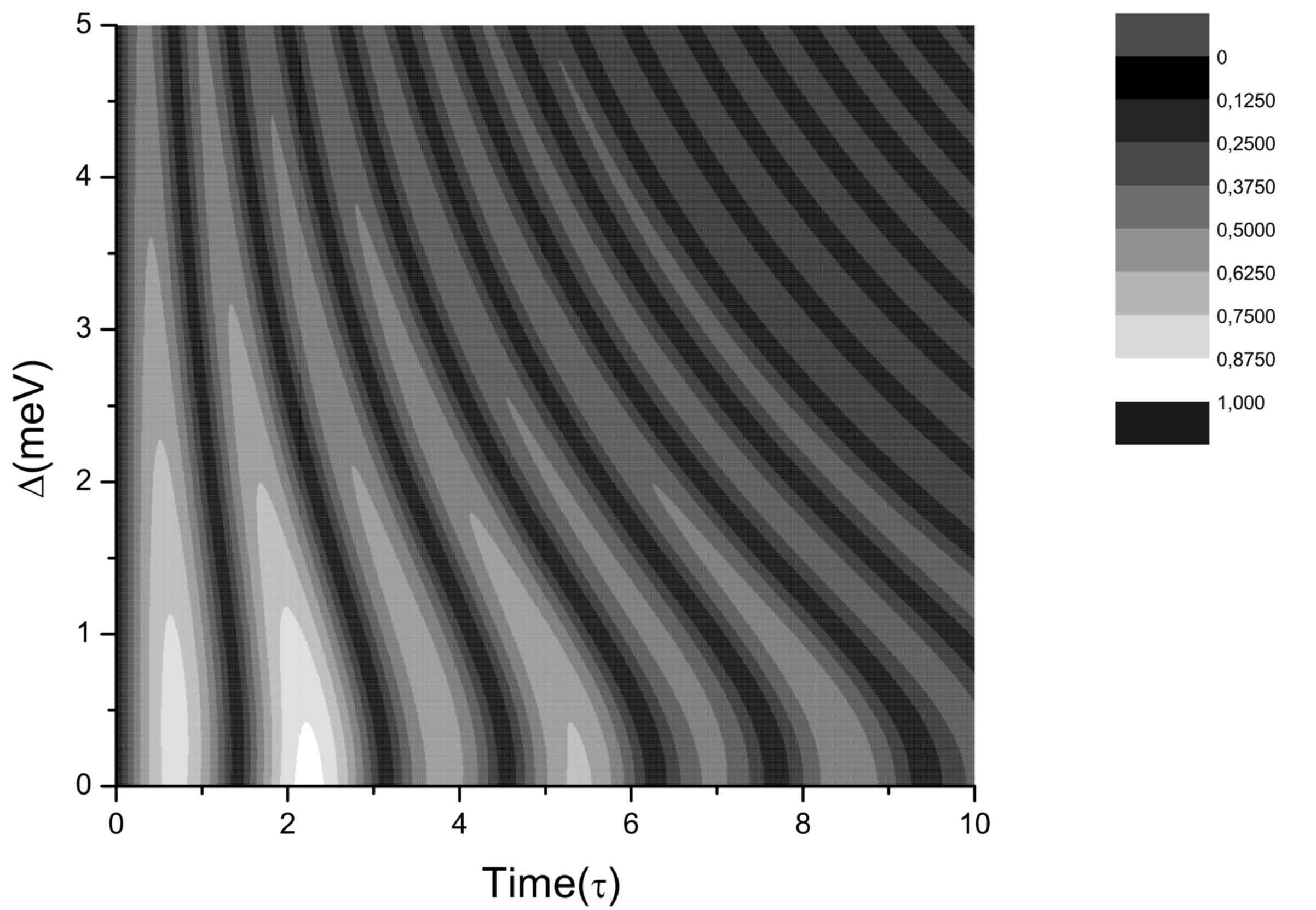}\\
  \caption{Concurrence as function of the detuning and the time, measured by the cross-coherence function in the strong coupling regime. With parameters $\alpha_0=0.01$, $\theta=\pi/2$, $g= 1$ meV,  $\omega=1$ eV, $\kappa=0.01$ meV, $\gamma=0.1$  meV and $\Delta=5$  meV.}\label{Cross_Concurrence}
  \end{center}
\end{figure}

\begin{figure}
  \begin{center}
  \includegraphics[width=12.5 cm]{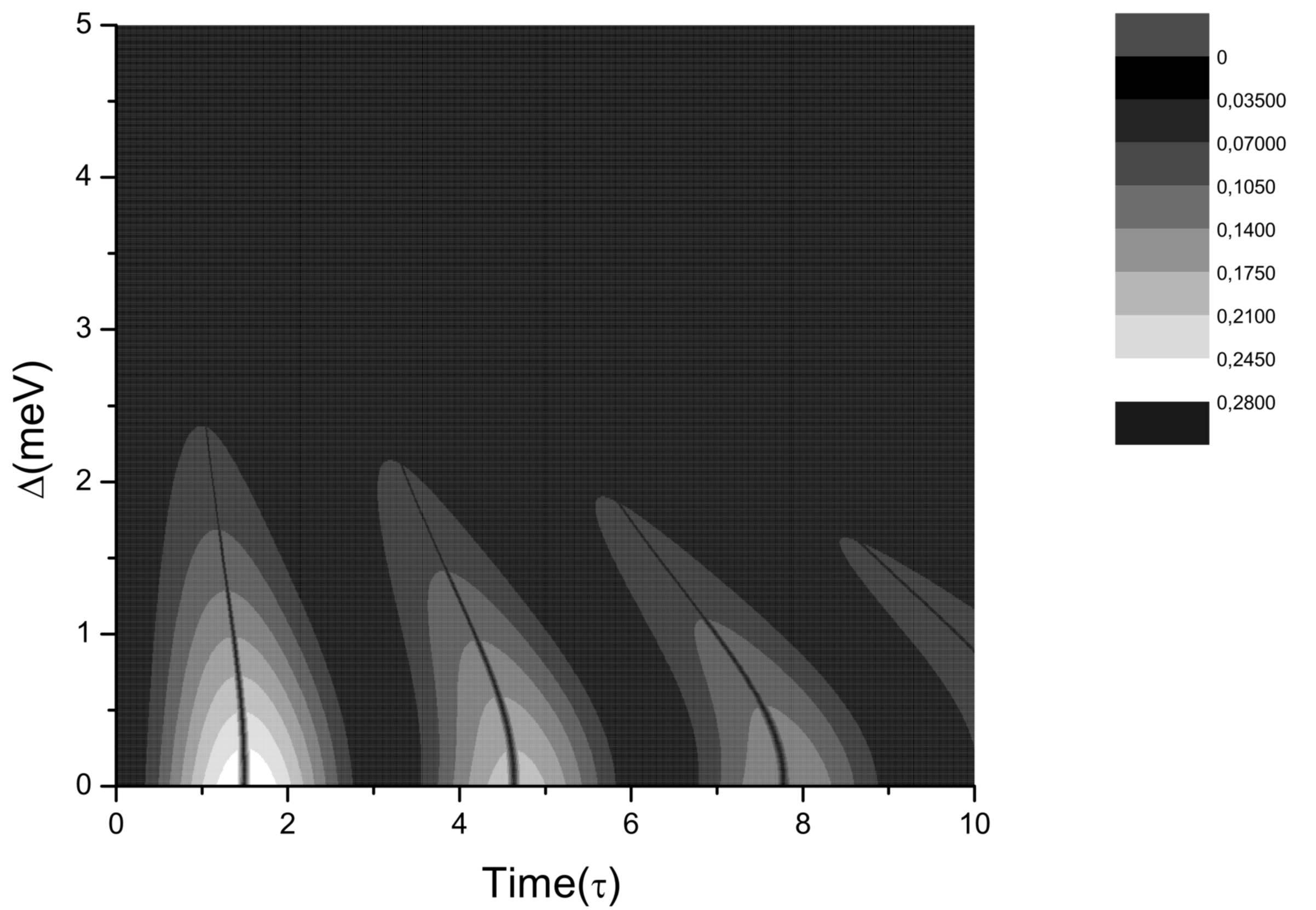}\\
  \caption{Error in the concurrence as measured by the cross-coherence function in the strong coupling regime. With parameters $\alpha_0=0.01$, $\theta=\pi/2$, $g= 1$ meV,  $\omega=1$ eV, $\kappa=0.01$ meV, $\gamma=0.1$ meV and $\Delta=5$ meV.}
  \end{center}
  \label{Cross_Concurrence_Err}
\end{figure}

A remarkable benefit of the factorization of the state of the system proposed in the equation (\ref{ansatz}). Is that the quantum correlations are completely encoded in the displaced ``backward" state of the system  $\tilde{\rho}(t')$. Here our aim is study the time evolution of the exciton-photon entanglement as measured by the Wooters Concurrence \cite{Wootters1998a}. That for a displaced state of the system ordered as in the equation (\ref{eq:Rhotilde}). Can be calculated in the following way. In first place, we define the spin-flipped state corresponding to $\tilde \rho$, $\tilde{\rho}_{SF}=(\sigma_y\otimes\sigma_y)\tilde{\rho}^{\star}(\sigma_y\otimes\sigma_y)$ where $\sigma_y$ is a paulimatrix, then we calculated the eigenvalues of the matrix $R=\sqrt{\sqrt{\tilde{\rho}} \tilde{\rho}_{SF} \sqrt{\tilde{\rho}}}$, thus the concurrence is defined as $C(\tilde{\rho})=max\{0,\lambda_1-\lambda_2-\lambda_3-\lambda_4\}$, where the $\lambda_i$ are the eigenvalues in decresing order. in terms of our displaced matrix elements, the concurrence reduces to
\begin{eqnarray} \fl
C(t) & = \left \vert
 \sqrt{\tilde{\rho}^{01}_{10}(t){\tilde{\rho}}^{10}_{01}(t)}
 +\sqrt{\tilde{\rho}^{01}_{01}(t) {\tilde{\rho}}^{10}_{10}(t)}
\right \vert
 -\left\vert
 \sqrt{{\tilde{\rho}}^{01}_{10}(t)\tilde{\rho}^{10}_{01}(t)}
 -\sqrt{{\tilde{\rho}}^{01}_{01}(t) \tilde{\rho}^{10}_{10}(t)}
\right\vert
\nonumber \\
\fl & = 2\sqrt{\tilde{\rho}^{01}_{10}(t)\tilde{\rho}^{10}_{01}(t)}
= 2 \vert \tilde{\rho}^{10}_{01}(t)\vert. \label{eq:correlations_impl} \\
\label{eq:free_concurrence}\fl C(t)&=2\left\|\frac{i \sqrt{2} g \sin ^2(\theta ) \sin \left(\frac{F1
   t}{\sqrt{2}}\right) e^{- (\gamma +\kappa )t}}{F1} \right\|,
\end{eqnarray}
where  we has defined the following set of auxiliary constants that are functions of the parameters of the system
\begin{eqnarray}
\fl F1= & \sqrt{4 g^2-\gamma ^2+\Delta ^2-\kappa ^2+ F2 +2 \gamma  \kappa }\\
\fl F2= & \sqrt{16 g^4-8 g^2 (\gamma -\Delta -\kappa ) (\gamma +\Delta -\kappa
   )+(\gamma -\kappa )^2+\Delta ^2}.
\end{eqnarray}
We now will make some remarks about the units we using in all the graphics in the paper. The energy  measured in mili-electronvolts and the time measured in units of $\frac{\hbar}{g},$ taking $g=1$  meV and $\hbar=1$. Thus one unit in time corresponds approximately to $\tau=6.58\times 10^{2}$ fs. The expression (\ref{eq:free_concurrence}) shows that the concurrence in the strong coupling regime, is an oscillatory function of time. That is slightly damped by effects of the decoherence as the times goes to infinity presenting regularly spaced in time collapses and revivals. We present a plot of  the expression for the concurrence (\ref{eq:free_concurrence}) in the figure (\ref{ODR}).
\clearpage
 Our main goal in this paper, is to show that an observable associated with Wotters Concurrence can be found in cross correlation experiments as it  has been proposed previously for a one-atom Laser in \cite{Kilin2007}, where the cross correlation function was being measured, multiplying in a correlator, the heterodyne  photocurrent produced by the photons in the cavity, by the heterodyne photocurrent produced by the photons generated by spontaneous emission in the matter emitter. For this purpose, now we will study the time resolved behavior of the quantum cross correlation function spectra, showing that near to a specific frequency that we will call  $\omega_{_{F}}^{(0)}$ the matter-field entanglement can be measured directly from the time resolved  cross correlation function spectra. We calculate the spectra related to the  first order cross correlation function $G^{(1)}_{_{Cr_2}}(t,\tau)$ via the real part of the Fourier transform of this function as it has been defined in the equation (\ref{eq:Kinchin}),  our interest in the quantum  cross correlation is due to the fact that this function depends explicitly on the coherence  that measure  the entanglement, as it can be seen in the   (\ref{eq:correlations}) and was proved in the equation (\ref{eq:correlations_impl}). The result for the spectra calculation is
\begin{equation}\label{eq:Cross_coherence2}
\fl I_{Cr_2}(t,\omega_{_{F}})=\frac{1}{\pi} \mathfrak{Re}\left\{\frac{4 i (\rho^{01}_{01}(t) g+\rho^{10}_{01}(t)(\Omega-\omega_{_{F}}-i \kappa  ))}{-4 g^2+(2\omega_f-\omega-\Omega+i(\gamma+\kappa))^2+(\gamma -\kappa -i \Delta )^2}\right\},
\end{equation}
where the information about the coherence of the density matrix $\rho_{01}^{10}(t)$, that determines the matter-field concurrence of the equation (\ref{eq:free_concurrence}) is present. Even more, comparing the equation (\ref{eq:Cross_coherence2}) with the expression for the light contribution to the photoluminescence spectra (\ref{eq:Light_spectra}), just in the case of the quantum cross correlation function, the coefficient that multiply the coherence $\rho_{01}^{10}(t)$ is a function of $\omega_{_F}$ that has a local extremal value for a real valued observational frequency. Due to this fact, in experiments just the quantum cross correlation function enable us to extract information about the coherence that determines the matter-field entanglement. With the aim to obtain an explicit expression of the contribution to the cross correlation spectra of  $I_{Cr_2}(t,\omega_{_{F}})$, we rationalize the equation (\ref{eq:Cross_coherence2}) that enable us to write $I_{Cr_2}(t,\omega_{_{F}})= A_P(\omega_{_{F}})\rho^{01}_{01}(t) +A_C(\omega_{_{F}})\rho^{10}_{01}(t)$, where we have defined the amplitude of the population and of the coherence in the following way
{\small\begin{eqnarray}
\fl A_C(\omega_{_{F}})= \frac{g^2 (\omega_{_{F}}-\Omega )+(\omega -\omega_{_{F}}) \left((\omega_{_{F}}-\Omega )^2+\kappa
   ^2\right)}{\pi  \left(g^4+2 g^2 (\gamma  \kappa -(\omega_{_{F}}-\omega )
   (\omega_{_{F}}-\Omega ))+\left((\omega_{_{F}}-\omega )^2+\gamma ^2\right) \left((\omega_{_{F}}-\Omega
   )^2+\kappa ^2\right)\right)}\\
\fl A_P(\omega_{_{F}})=\frac{g (-\omega_{_{F}} (\gamma +\kappa )+\gamma  \Omega +\kappa  \omega )}{\pi
   \left(g^4+2 g^2 (\gamma  \kappa -(\omega_{_{F}}-\omega ) (\omega_{_{F}}-\Omega
   ))+\left((\omega_{_{F}}-\omega )^2+\gamma ^2\right) \left((\omega_{_{F}}-\Omega )^2+\kappa
   ^2\right)\right)}.
\end{eqnarray}}
Thus, the simpler guess for a criterion to determine the frequency where the coherence $\rho^{01}_{10}(t)$ could be measured from the quantum cross correlation spectra in an experiment as the proposed in \cite{Kilin2007}, is obtained maximizing the numerator of $A_C(\omega_{_{F}})$ with respect to $\omega_{_{F}}$, the contribution of this amplitude is maximum at
\begin{equation}
\omega_{_{F}}^{(0)}=\frac{1}{3} \left(-\sqrt{3 g^2-3 \kappa ^2+\omega ^2-2 \omega  \Omega
   +\Omega ^2}+\omega +2 \Omega \right),
\end{equation}
that is always a real frequency, and therefor constitutes a good observational frequency. To illustrate better how the criterion works, we plot the absolute value of  the two amplitude as function of $\omega_{_{F}}$ in, and out of resonance in figure (\ref{Expl_criteria}), showing with an arrow the zone where the observational frequency $\omega_{_{F}}^{(0)}$ is located.
  With the aim of study how the criterion works for a broad region of the parameters space, we have calculated the concurrence in terms of the cross coherence function contribution to the quantum cross correlation function spectra as $\left|\frac{2 I_{Cr_2}(t,\omega_{_{F}}^{(0)})}{A_C(\omega_{_{F}}^{(0)})}\right|$, we present this result in a plot in the figure (\ref{Cross_Concurrence}), we also have  calculated the error in the concurrence determination by this approach as $\left|C(t)- \left|\frac{2 I_{Cr_2}(t,\omega_{_{F}}^{(0)})}{A_C(\omega_{_{F}}^{(0)})}\right|\right|$, a plot in the figure (\ref{Cross_Concurrence_Err}) show this result for our oversimplified model, the guess of measure the concurrence directly from the cross correlation function spectra works relatively well in a broad region of the relevant parameters space. It is remarkable that the result in figure (\ref{Cross_Concurrence_Err}) shows that the optimal behavior of the present proposal for the measure of the concurrence, works very well for detuning beyond $2.5$ meV. It is important to remark that in the same spirit that we determine  the criterion to measure the concurrence from the crossed coherence function, the maximization procedure can be done to measure other observable quantities from the contributions to the quantum cross correlation, and photoluminescence spectra.

\section{Discussion}
\begin{figure}[h!]
\begin{center}
\includegraphics[width=14 cm]{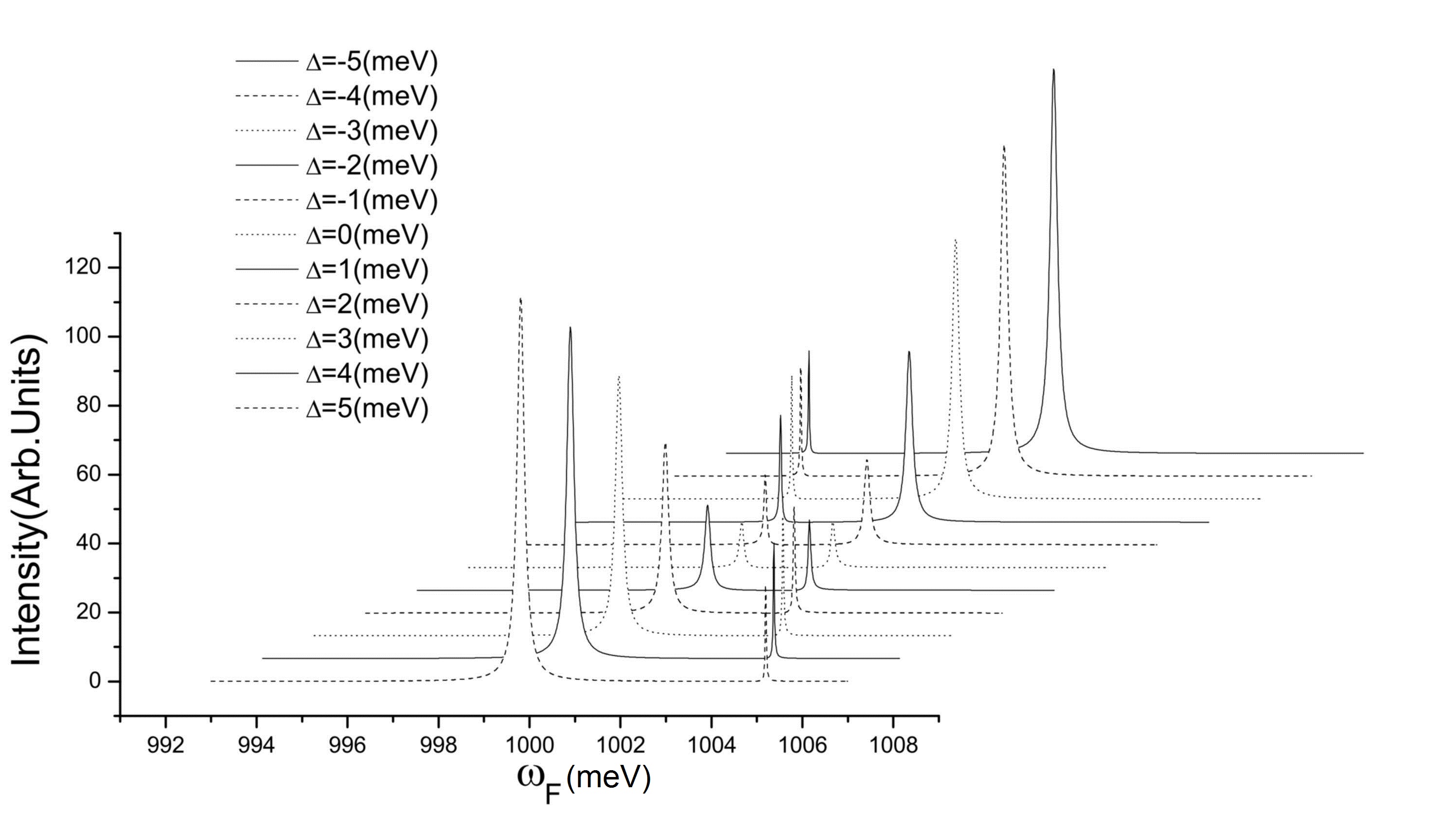}\\
\caption{Photoluminescence spectra as function of the field-matter detuning parameter, the left branch corresponds to the polariton state $\ket{\chi_0}$ and the right branch correspond to the polariton state $\ket{\chi_1}$, see the appendix 1. Using the parameter values $\alpha_0=0.01$, $g= 1$ meV,  $\theta=\pi/2$, $\omega=1$ eV, $\kappa=0.01$ meV, and $\gamma=0.1$ meV.}\label{AC}
\end{center}
\end{figure}

\begin{figure}[h!]
  \begin{center}
  \includegraphics[width=12.5 cm]{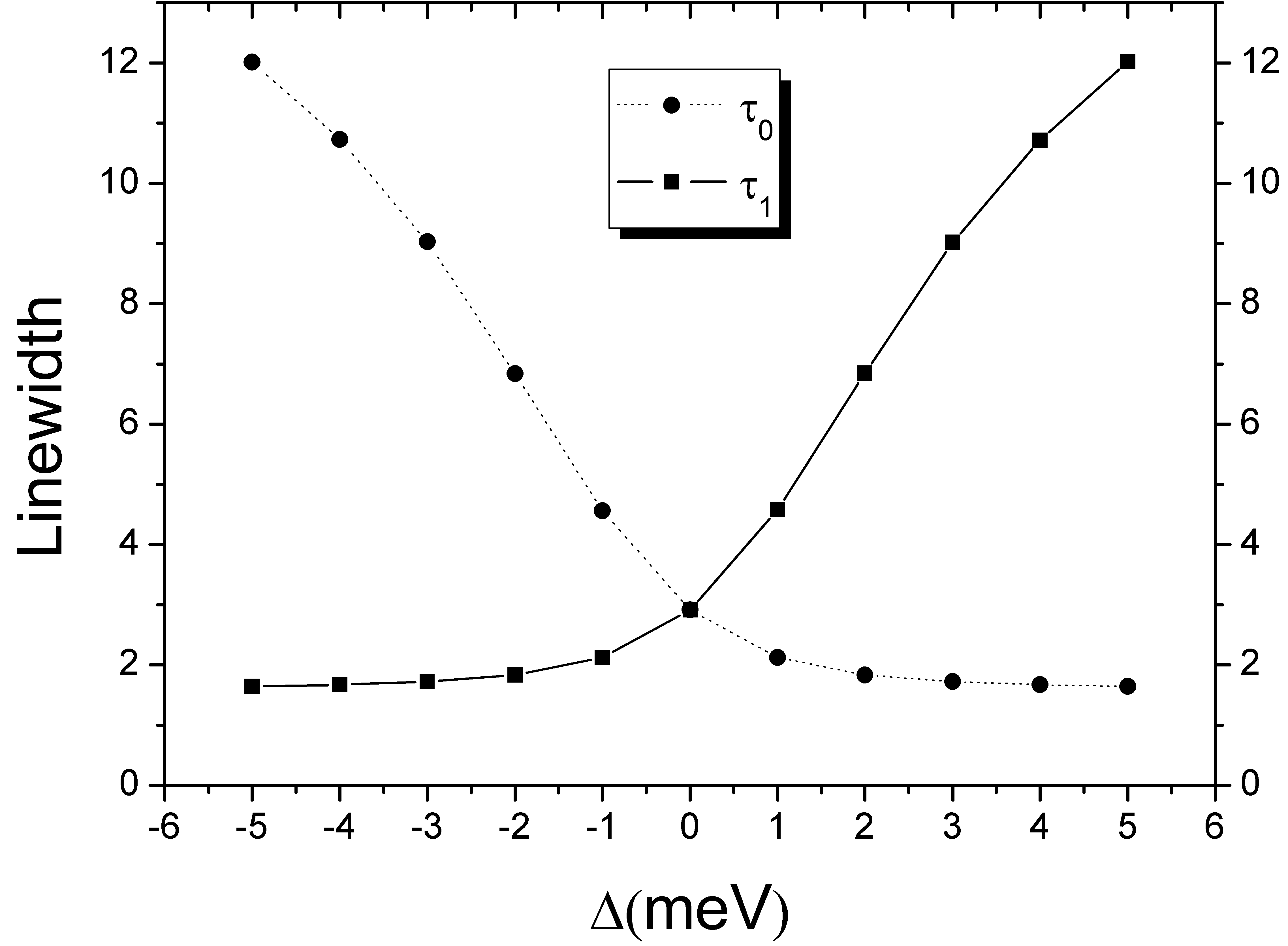}\\
  \end{center}
  \caption{In squares the linewidth as function of time for polariton branch corresponding to the state $\ket{\chi_0}$;  in circles the same for the polariton branch corresponding the state $\ket{\chi_1}$. With parameters $\alpha_0=0.01$, $\theta=\pi/2$, $g= 1$ meV,  $\omega=1$ eV, $\kappa=0.01$ meV, $\gamma=0.1$ meV and $\Delta=5$ meV.}\label{LT}
\end{figure}

\begin{figure}[h!]
  \begin{center}
  \includegraphics[width=12.5 cm]{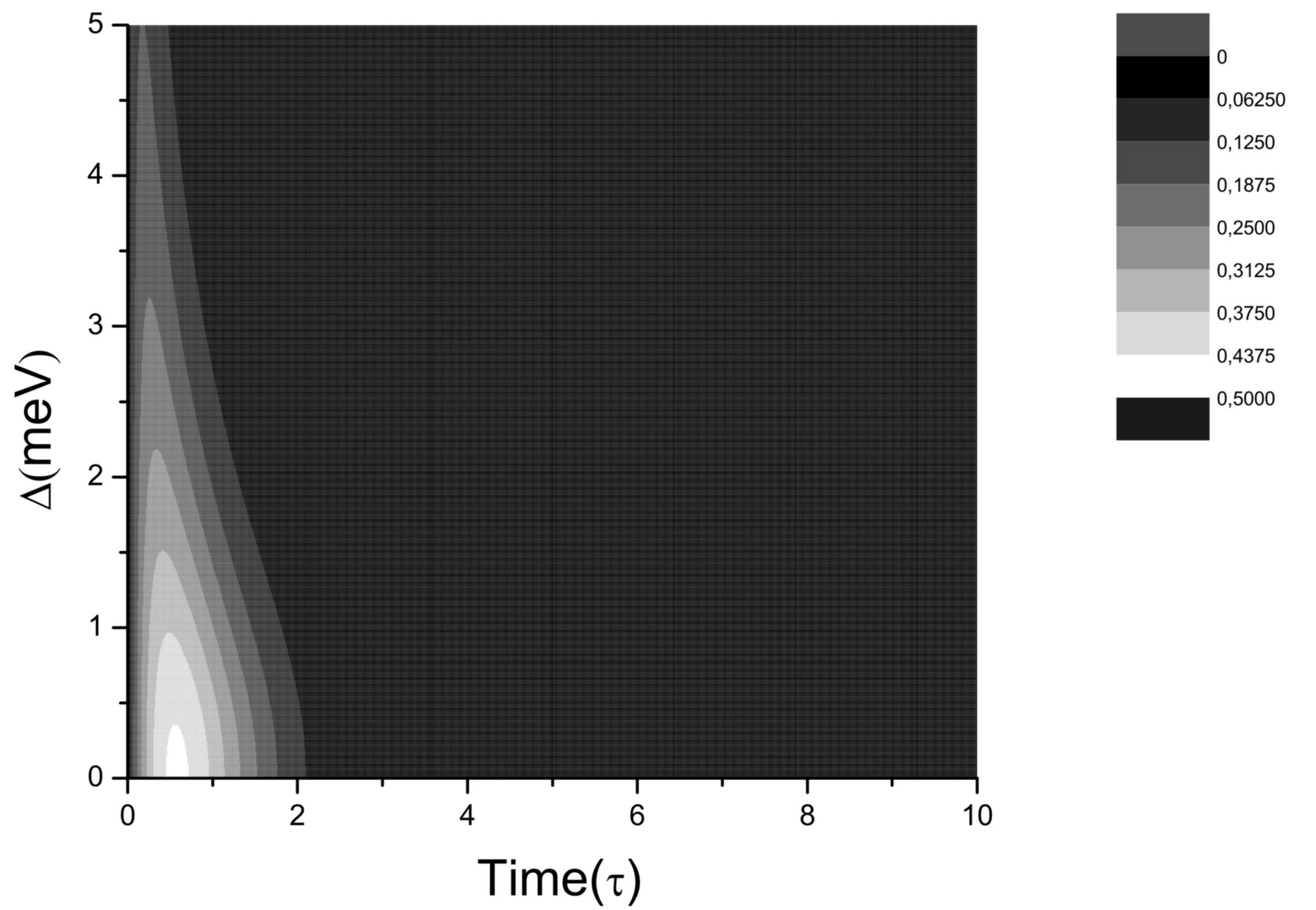}\\
  \end{center}
  \caption{Concurrence as function of time and detuning in the weak coupling regime, in this graphic the time scales in units of $\tau\approx6.58\times 10^{2}$ fs, with $g= 1$ meV, $\kappa=0.01$ meV and $\theta=\pi/2;$ we choose for this graphic the parameter $\gamma=\sqrt{\frac{1}{8 g^2-1}} \sqrt{16 g^4+8 g^2 \Delta ^2+\Delta ^2}+\kappa$ that fulfill the condition established in the equation (\ref{cond_WC}) for the weak coupling regime of the dynamics.}\label{CSCR}
\end{figure}

\begin{figure}[h!h!t!]
\begin{center}$
\begin{array}{cc}
\includegraphics[width=7.5cm]{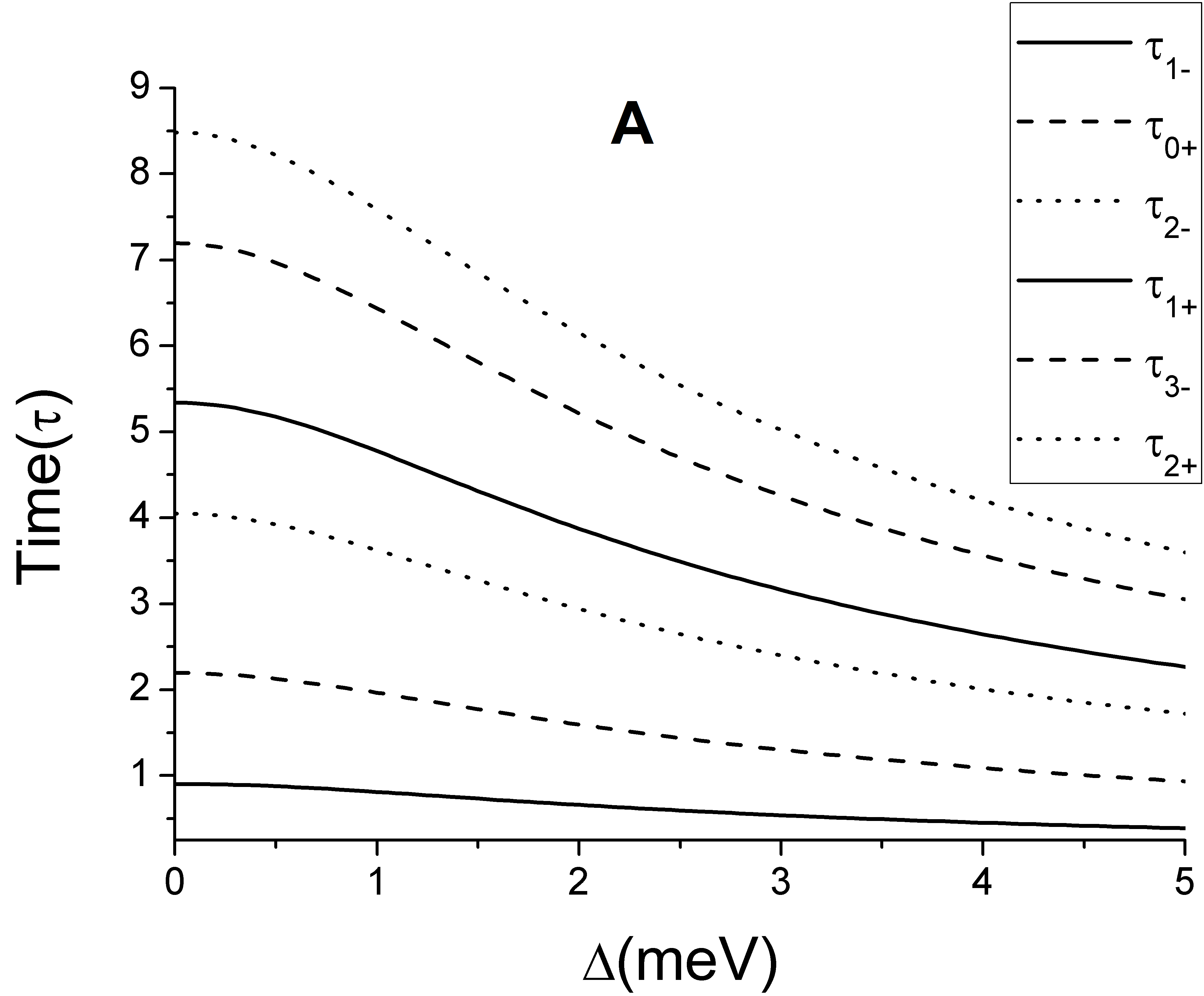} &
\includegraphics[width=7.5cm]{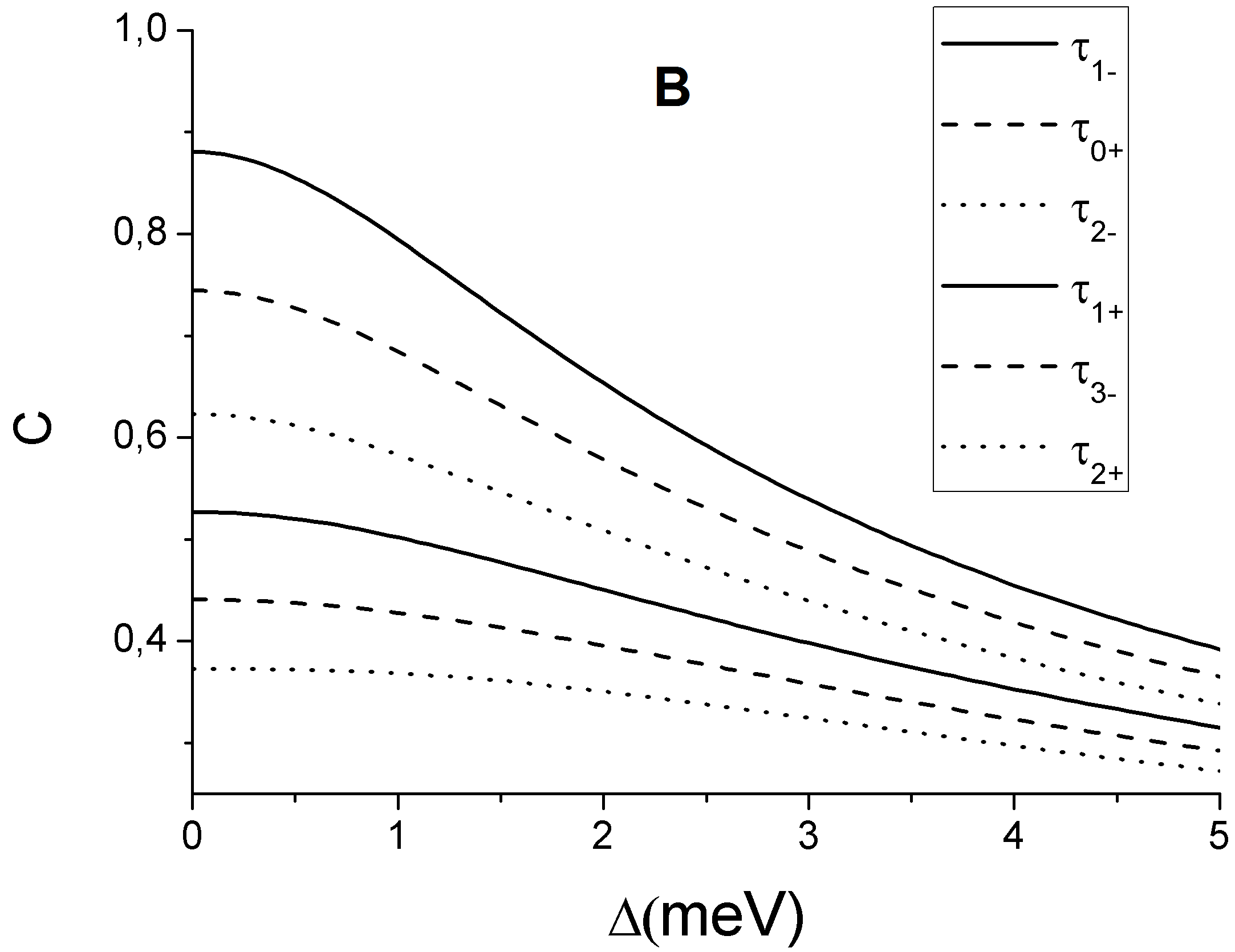}
\end{array}$
\end{center}
\caption{In A, the maximum concurrence times as a function of the detuning, where the time scales in units of $\tau\approx6.58\times 10^{2}$ fs, and the behavior of the value of the concurrence at this times in B, using the parameter values $g= 1$ meV,  $\theta=\pi/2$, $\omega=1$ eV, $\kappa=0.01$ meV, and $\gamma=0.1$ meV.}\label{Max_Conc_times}
\end{figure}

Based in the expression for the concurrence (\ref{eq:free_concurrence}) we identify the  dynamical coupling regimes, analyzing the four values of $\kappa$ that cancels the constant $F1$, it is
\begin{equation}
\kappa=\frac{1}{2} \left(2 \gamma \pm\sqrt{2} \sqrt{1+2 \Delta ^2\pm\sqrt{8 \Delta
   ^2+1}}\right),
\end{equation}
the former result remain valid after the roles  of the parameters $\gamma$ and $\kappa$ are exchanged. The system is in critical damping regime when $F1\rightarrow0$, in this regime the concurrence takes the value $C(t)=2\left \vert g t \sin ^2(\theta ) e^{-(\gamma +\kappa )t} \right\vert.$ In order that the system be in the strong coupling dynamical regime, defined as the regime of the dynamics in which the eigenstates of the system are entangled matter-light states. $F1$ must be a real number, it is fulfilled in any of the following four possible configurations of parameters
\begin{eqnarray}\label{Cond_SC}
\Delta \leq -1\land \kappa <\frac{1}{2} \left(2 \gamma +\sqrt{2}
   \sqrt{2 \Delta ^2+\sqrt{8 \Delta ^2+1}+1}\right)\\
-1<\Delta <0\land \kappa <\frac{1}{2} \left(2 \gamma +\sqrt{2} \sqrt{2
   \Delta ^2-\sqrt{8 \Delta ^2+1}+1}\right)\\
   0<\Delta <1\land \kappa <\frac{1}{2} \left(2 \gamma +\sqrt{2} \sqrt{2
   \Delta ^2-\sqrt{8 \Delta ^2+1}+1}\right)\\
   \Delta \geq 1\land \kappa <\frac{1}{2} \left(2 \gamma +\sqrt{2} \sqrt{2
   \Delta ^2+\sqrt{8 \Delta ^2+1}+1}\right).
\end{eqnarray}
Again, the former result remain valid after the roles  of the parameters $\gamma$ and $\kappa$ are exchanged. It is remarkable that, in this situation the dynamical regime does not depend on the matter-field coupling strength $g$.  By the other hand, the strong coupling regime of the dynamics is often established through  the appearing of an anticrossing as function of the detuning in the photoluminescence spectra. With this in view, we will study the stimulated emission process contribution to the time integrated photoluminescence spectra that is defined in terms of the first order correlation function  $G^{(1)}_{_{C}}(t,\tau)$. Introducing the constant  ${\tilde g}=\sqrt{-4 g^2+(\gamma -i \Delta -\kappa )^2}$, we find the following result for this contribution to the photoluminescence spectra
\begin{equation}
I_{C}(t,\omega_{_{F}})=\frac{1}{\pi} \mathfrak{Re}\left\{-\frac{4 i (\rho^{01}_{10}(t) g+\rho^{10}_{10}(t) (-\omega_{_{F}}-i \kappa +\Omega ))}{{\tilde g}^2+(2 \omega_{_{F}}+i (\gamma +\kappa +i (\omega +\Omega )))^2}\right\}.\label{eq:Light_spectra}
\end{equation}
To obtain the contribution to the time integrated photoluminescence spectra, the former result must be integrated in time as
\begin{equation}
I_{C}(\omega_{_{F}})=\lim_{T\rightarrow\infty}\int_{0}^{T}I_{C}(t,\omega_{_{F}})dt.
\end{equation}
We present a plot of $I_{C}(\omega_{_{F}})$ for some values in the detuning at the figure (\ref{AC}). The typically accepted, signature that the system is in strong coupling regime, is comprised in the apparition of  an anti-crossing in the photoluminescence spectra, near to resonance. It can be seen in the figure (\ref{AC}), where each branch in the spectra corresponds to a polaritonic excitation in the system. Where our criterion for the strong coupling (\ref{Cond_SC}) has been used, in order to show consistency with the typically accepted criterion e.g. \cite{StrongCoupling}.
Interpolating the photoluminescence spectra profiles, for each value of the field-matter detuning parameter, with a Lorentzian profile. We obtain the linewidth for each polariton branch, the guess of Lorentzian interpolation is in good approximation for our model of coupled oscillators in the ultralow density of excitons regime. The linewidth of the polariton branches calculated within the present setup can be seen in the figure (\ref{LT}). The crossing in the linewidth at resonance that can be observed in the figure (\ref{LT}, constitute another signature of the strong coupling regime of the dynamics that is typically accepted. It is important to remark here, that the values in the parameters that we use corresponds to microcavities with a lower quality factor than the used in the reference \cite{StrongCoupling}, fact that can imply a strong discrepancy in the states linewidth when compared with the result reported there in.
By means of an analogous reasoning, that the one that conduce us to the result (\ref{Cond_SC}), in order to have, the system be in the weak coupling dynamical regime, defined as the dynamical regime in which the eigenstates of the system are very closer to the bare matter-field states, $F1$ must be pure imaginary number. This condition is accomplished when
\begin{equation}
\label{cond_WC} \gamma \geq \sqrt{\frac{1}{8 g^2-1}} \sqrt{16 g^4+8 g^2 \Delta ^2+\Delta ^2}+\kappa,
\end{equation}
for all the values of the detuning parameter, again the former result remain valid after the roles  of the parameters $\gamma$ and $\kappa$ are exchanged but this time the result depend on the value of the field matter coupling constant $g$. In contrast with the behavior of the concurrence in the strong coupling regime, in the present situation, the concurrence is not more  periodic in time, reaching a maximum value for finite time, and tends asymptotically to zero as time tends to infinity, we plot this result as a function of the detuning and the time in the figure (\ref{CSCR}).
Maximizing the expression for the concurrence in (\ref{eq:free_concurrence}) we find the following expression for the times at which the concurrence is a maximum
\begin{eqnarray}
\fl \tau_{m\pm}=\frac{\sqrt{2} \left(2 \tan ^{-1}\left(\frac{\pm\sqrt{F1^2+2 \gamma
   ^2+4 \gamma  \kappa +2 \kappa ^2}-\sqrt{2} \gamma -\sqrt{2} \kappa
   }{F1}\right)+2 \pi  m\right)}{F1},
\end{eqnarray}
where $m$ run over the integer numbers and the maxima are ordered as $\tau_{1-},\tau_{0+},\tau_{2-},\tau_{1+},\dots,$ by the other hand, the concurrence take its minima values for times
\begin{equation}
{\tilde\tau}_{m}=\frac{\sqrt{2} (2 \pi  m+\pi )}{F1}.
\end{equation}
It will result enlightening from a qualitative point of view, study the behavior of the maximum concurrence times, and the behavior of the concurrence at these times, with this aim in view, we plot thess quantities in the figure (\ref{Max_Conc_times}).
Neither the expression for the concurrence in the equation (\ref{eq:free_concurrence}), neither the maxima or minima concurrence times depends on $\Omega$ and $\omega$, depends rather on its difference $\Delta$. Then in the case of coherent photons pumping, the concurrence evolves too as in the equation (\ref{eq:free_concurrence}).

\begin{figure}[h!t!]
  \begin{center}
  \includegraphics[width=11.5 cm]{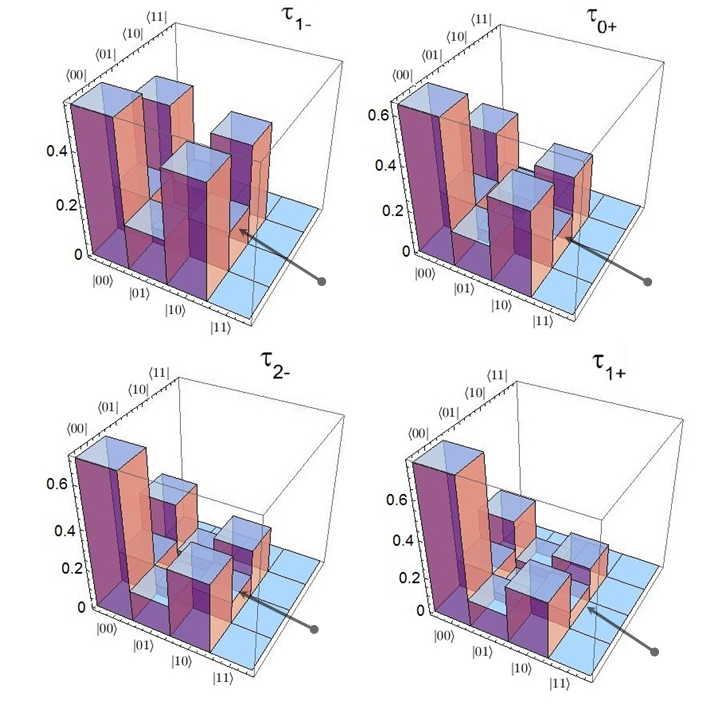}\\
  \end{center}
  \caption{[Color online] Tomography of the density matrix for the first four maximum concurrence times. With the arrow signaling the position of the coherence $\rho_{01}^{10}(\tau)$ in modulus for each time, the parameters we used was $\alpha_0=0.01$, $\Delta=5$ meV, $\theta=\pi/2$, $g= 1$ meV,  $\omega=1$ eV, $\kappa=0.01$ meV, and $\gamma=0.1$ meV.}\label{Tomography}
\end{figure}
\clearpage
\section{Concluding remarks}
We study the matter-field entanglement in a semiconductor microcavity. Using a model  encompassed by two driven  damped and  coupled oscillators, with a detuning $\Delta$ between them. Where  the first one is for the photons, and the second one is for the excitons in the microcavity. This Model represents accurately, a semiconductor microcavity, for experiments where the condition of ultra low density of excitons is fulfilled. It is, $N a_{_B}^{d}\ll 1$ where $N$ is the number of excitons, $a_{_B}$ is the Bohr radius by each  exciton and $d$ is the dimensionality of the heterostructure. We report a closed analytical expression that quantify the exciton-photon entanglement in terms of $\gamma,$ $\kappa$ and $\Delta$, using the Wotters Concurrence (\ref{eq:free_concurrence}). Valid for  experimental setups  with or without coherent pumping of photons. Showing that accurated information about the dynamics of the concurrence can be measured indirectly. V\'ia the cross correlation function (\ref{eq:Cross_coherence2} ). That can be measured multiplying in a correlator device, the heterodyne photocurrents produced by the photons in the microcavity, and the excitons recombination. If both are measured in a time resolved way, within a narrow window in frequency, centered at the observational frequency $\omega_{_F}^{(0)}$.
\section{Acknowledgments}
 I thank to Mazda Foundation for the Arts and the Science for financial support, also to K. M. Fonseca-Romero and Herbert Vinck-Posada for useful and stimulating discussions.
\section*{Appendix 1: Bare system eigenstates concurrence}\label{ap:Bare_eigen}
Expanding the Hamiltonian (\ref{eq:Ham_Free}) in the base with zero and one excitation,  the Hamiltonian of the system present two eigenstates with nontrivial dynamics that are
\begin{eqnarray}
\ket{\chi_0}=\frac{g\ket{01}-\frac{1}{2}(-\Delta+\sqrt{\Delta^2+4g^2})\ket{10}}{\sqrt{2g^2+\frac{1}{2}\Delta^2-\Delta\sqrt{g^2+\frac{\Delta^2}{4}}}}\\
\ket{\chi_1}=\frac{g\ket{01}-\frac{1}{2}(-\Delta-\sqrt{\Delta^2+4g^2})\ket{10}}{\sqrt{2g^2+\frac{1}{2}\Delta^2-\Delta\sqrt{g^2+\frac{\Delta^2}{4}}}}
\end{eqnarray}
often this states are called dressed states or polaritonic excitations of the system and each one have associated an eigenenergy
\begin{equation}
E_{0/1}=\frac{1}{2}(2\omega+\Delta\pm\sqrt{\Delta^2+4g^2})
\end{equation}
that define the two polaritonic branches of the  photoluminescence spectra of the system. As it can be seen, the two states are non-separable or entangled, with the same Wootters concurrence --see the reference \cite{Wootters1998a}-- for both dressed states and determines the maximum matter-field entanglement available within the system, and is equal to
\begin{equation}\label{eq:dressed_concurrence}
C_D(\Delta)=2 \sqrt{\frac{g^2}{4 g^2+\Delta ^2}}.
\end{equation}
\section*{Appendix 2: Coherent amplitude time evolution}\label{ap:Cohe_ampl}
In the case that the pumping is not present, the dynamics of the coherent amplitude is described by
\begin{eqnarray}
\label{eq:dDisplacementdt}  \frac{d}{dt} \left(
\begin{array}{c}
\alpha \\ \beta
\end{array}
\right ) = \left(
\begin{array}{rc}
-(\gamma+i\omega) & -i g \\
-i g  & -(\kappa + i\Omega)
\end{array}
\right ) \left(
\begin{array}{c}
\alpha \\ \beta
\end{array}
\right).
\end{eqnarray}
Solving the evolution equations for the coherent amplitude in the equation (\ref{eq:dDisplacementdt}), we find that they evolve in time in the following way
\begin{eqnarray}
\fl \label{eq:alpha} \alpha(t) =\frac{e^{-\frac{1}{2} t (i
\tilde{g}+\gamma +\kappa +i (\omega +\Omega))} \alpha_0
\left(\tilde{g} \cos \left(\frac{ \tilde{g} t}{2}\right)+(\gamma
-\kappa -i \omega +i\Omega ) \sin \left(\frac{\tilde{g}
t}{2}\right)\right)}{\tilde{g}},
\\
\label{eq:beta}\fl \beta(t)  = \frac{2 e^{-\frac{1}{2} t (\gamma
+\kappa +i (\omega +\Omega ))} g \alpha_0 \sin \left(\frac{\tilde{g}
t}{2}\right)}{\tilde{g}},
\end{eqnarray}
where we have defined  $\tilde{g}=\sqrt{(\kappa-\gamma -i \Delta)^2-4 g^2}$ with $\Delta=\Omega -\omega $ often called the detuning parameter. In the case of pumping is present, the coherence amplitude evolves according to
\begin{eqnarray}
\label{eq:dDisplacementdt_pump}\fl \left(
         \begin{array}{c}
           \dot{\alpha}(t) \\
           \dot{\beta}(t) \\
         \end{array}
       \right)=\left(
\begin{array}{cc}
 -\gamma -i \delta  & -i g \\
 -i g & -\kappa -i (\delta +\Delta )
\end{array}
\right)\left(
         \begin{array}{c}
           \alpha(t) \\
           \beta(t) \\
         \end{array}
       \right)-\left(
                 \begin{array}{c}
                   i \mathcal{F} \\
                   0 \\
                 \end{array}
               \right)
\end{eqnarray}
that must be solved using the initial conditions in the equation (\ref{ec:initialystate1}), taking again $\alpha(0)=\alpha_0$ and $\beta(0)=0$. We find the following result for the evolution of the coherent amplitude
{\tiny \begin{eqnarray}
\fl \alpha(t)=\frac{2 e^{-\frac{1}{2} t (\gamma +2 i \delta +i \Delta +\kappa )}\left(-2e^{\frac{1}{2}t (\gamma +2 i \delta +i \Delta +\kappa )} F B_{1}-2 B_{2} \cos\left(\frac{\tilde{g} t}{2}\right)+2 i\tilde{g} B_{3} \sin\left(\frac{\tilde{g}   t}{2}\right)\right)}{\left((-\gamma +i \Delta +\kappa)^2-4g^2\right) \left(\tilde{g}^2+(\gamma +i (2 \delta +\Delta )+\kappa)^2\right)} \\ \nonumber
\fl B_{1}=\left(4g^2-((\kappa-\gamma)+i \Delta )^2\right)((\delta+\Delta) -i \kappa)\\ \nonumber
\fl B_{2}=\left(4g^2-((\kappa-\gamma) +i \Delta )^2\right) \left(g^2 \alpha_{0}-(F+\alpha_{0}(\delta -i \gamma )) ((\delta+\Delta) -i\kappa )\right)\\ \nonumber
\fl B_{3}=\left(\alpha_{0} \left((\gamma +i \delta ) ((\delta+\Delta) -i \kappa )-i g^2\right) (\gamma -i \Delta -\kappa )+F\left(2 g^2+(\gamma -i \Delta -\kappa ) (i (\delta+\Delta) +\kappa)\right)\right)\\
\fl \beta(t)=\frac{4 e^{-\frac{1}{2} t (\gamma +i (2 \delta +\Delta )+\kappa )}g^2\left(e^{\frac{1}{2} t (\gamma +i (2 \delta +\Delta )+\kappa )}FA_{1}+F A_{2} \cos \left(\frac{\tilde{g} t}{2}\right)+i \tilde{g} A_{3}\sin\left(\frac{\tilde{g}t}{2}\right)\right)}{\left((-\gamma +i \Delta +\kappa )^2-4g^2\right) \left(\tilde{g}^2+(\gamma +i (2 \delta +\Delta )+\kappa )^2\right)}\\ \nonumber
\fl A_1=\left(4g^2-((\kappa-\gamma) +i \Delta )^2\right)\\ \nonumber
\fl A_{2}=\left(((\kappa-\gamma) +i\Delta)^2-4 g^2\right)\\ \nonumber
\fl A_{3}=\left(i F (\gamma +i (2\delta +\Delta )+\kappa )+2 \alpha_{0} \left(g^2+(\gamma +i \delta ) (i (\delta+ \Delta)+\kappa)\right)\right)\\ \nonumber
\fl \tilde{g}=\sqrt{4 g^2-(\gamma + (i\Delta -\kappa ))^2}.
\end{eqnarray}}
\section*{References}
%
\bibliographystyle{srt}

\end{document}